\documentclass[runningheads, 10pt]{llncs}

\usepackage{graphicx}
\usepackage[most]{tcolorbox}
\usepackage{xcolor}
\usepackage[font=scriptsize,labelfont={scriptsize,bf}]{subfig}

\usepackage{booktabs}
\usepackage{tabularx}

\usepackage{amstext}  
\usepackage{xspace}
\usepackage{amssymb}
\usepackage[overload]{empheq}

\usepackage{caption}
\usepackage{graphbox}
\usepackage[colorlinks=true,linkcolor=blue,breaklinks=True,citecolor=brown,urlcolor=blue]{hyperref}

\usepackage{colortbl}
\usepackage{cellspace}

\newcommand{\textbox}[1]{
    \noindent\fbox{%
        \parbox{0.98\columnwidth}{%
            #1
        }%
    }
}

\newcommand{\vit}{\texttt{ViT}\xspace}
\newcommand{\swin}{\texttt{Swin}\xspace}
\newcommand{\siamese}{\texttt{Siamese}\xspace}
\newcommand{\generator}{\texttt{Generator}\xspace}
\newcommand{\discriminator}{\texttt{Discriminator}\xspace}
\newcommand{\lstpar}[1]{{\footnotesize\textit{(#1)}}}

\newtcolorbox{cooltextbox}[1][]{%
    colback=black!5,
    colframe=black!5,
    notitle,
    sharp corners,  
    borderline west={0pt}{0pt}{red!80!black},
    enhanced,
    breakable,
    left=0pt,
    right=0pt,
    top=0pt,
    bottom=0pt
    }

\begin{document}
\title{Attacking logo-based phishing website detectors with adversarial perturbations}

\titlerunning{Attacking logo-based phishing detectors with adversarial perturbations}

\author{Jehyun Lee\inst{1} \and
Zhe Xin\inst{2} \and
Melanie Ng Pei See\inst{2} \and 
Kanav Sabharwal\inst{2} \and \\
Giovanni Apruzzese\inst{3} \and
Dinil Mon Divakaran\inst{2,4}
}

\authorrunning{J. Lee et al.}

\institute{Trustwave \and  National University of Singapore \and
Liechtenstein Business School, University of Liechtenstein \and 
Acronis Research
}

\maketitle 

\begin{abstract}

Recent times have witnessed the rise of anti-phishing schemes powered by deep learning (DL). In particular, logo-based phishing detectors rely on DL models from Computer Vision to identify logos of well-known brands on webpages, to detect malicious webpages that imitate a given brand. For instance, Siamese networks have demonstrated notable performance for these tasks, enabling the corresponding anti-phishing solutions to detect even ``zero-day'' phishing webpages.
In this work, we take the next step of studying the robustness of logo-based phishing detectors against adversarial ML attacks. We propose a novel attack leveraging generative adversarial perturbations to craft ``adversarial logos'' that, with no knowledge of phishing detection models, can successfully evade the detectors. We evaluate our attacks through: (i)~experiments on datasets containing real logos, to evaluate the robustness of state-of-the-art phishing detectors; and (ii)~user studies to gauge whether our adversarial logos can deceive human eyes. The results show that our proposed attack is capable of crafting perturbed logos subtle enough to evade various DL models---achieving an evasion rate of up to 95\%. Moreover, users are not able to spot significant differences between generated adversarial logos and original ones.

\end{abstract}

\section{Introduction}
\vspace{-0.2cm}

Phishing attacks are on the rise~\cite{APWG-report-2023}, and they represent a serious threat to both organizations and individuals alike. While there have been numerous  research efforts to counter this long-running security problem~\cite{garera2007framework,Google-content-phishing-2010,URLnet-2018,d-fence-2021}, a universal solution against phishing has yet to be found, as  new ways to lure unaware victims keep emerging~\cite{BitB-2022}.  We focus on the problem of detecting phishing \textit{websites}, which has witnessed 61\% increase in 2022~\cite{phishreport2022url}.

The first line of defense against phishing websites is represented by blocklists, which are nowadays leveraged at scale~\cite{kondracki2021catching}. Unfortunately, such rule-based countermeasures only work against the phishing entries in the blocklist, and attackers are well-aware of this (for a recent report, see~\cite{cofense2023report}). To protect users against evolving phishing websites, current anti-phishing schemes are now equipped with data-driven methods that detect malicious webpages by leveraging some heuristics~\cite{gsb}. In particular, the constant progress and successes of \textit{machine learning} (ML) algorithms in research~\cite{tian2018needle,canita+-2011} led to the integration of ML-based phishing detectors also in popular browsers~\cite{liang2016cracking}.

There are various ways in which ML is used to identify phishing websites, depending on the input analyzed by the ML model~\cite{divakaran2022phishing}: URL (e.g.,~\cite{verma2015character,URLnet-2018}), HTML contents (e.g.,~\cite{Google-content-phishing-2010,canita+-2011,phishing-NDSS-MADWeb-2020}), or visual representations (e.g.,~\cite{corona2017deltaphish,visualphishnet}) of a webpage. Detection methods based on visual analytics are now receiving much attention (e.g.,~\cite{corona2017deltaphish,logosense,visualphishnet,lin2021phishpedia,logomotive-2022,phishintention-usenix-sec-2022}), likely due to the tremendous advancements in deep learning (DL). In this work, we delve into the application of DL for \textit{logo-based} phishing website detection---a state-of-the-art approach\footnote{\textbf{Background:} in simple terms, logo-based phishing detection seeks to identify those (malicious) webpages that attempt to imitate a well-known brand. Intuitively, if a given webpage has the logo of a well-known brand (e.g., PayPal), but the domain does not correspond to the same brand (e.g., www.p4y-p4l.com), the webpage is classified as phishing. Though these approaches require maintenance of a database of logos for brands, such a task is not impractical given that the number of brands targeted by attackers is typically small ($\approx200$)~\cite{visualphishnet,phishing-pandemic-2020,lin2021phishpedia}.} that is~\lstpar{i}~considered in recent researches (e.g.,~\cite{logosense,logomotive-2022,lin2021phishpedia,phishintention-usenix-sec-2022}), and \lstpar{ii}~deployed in practice~\cite{apruzzese2023position}.

In logo-based detection, the first task is to extract the logo(s) from a webpage (typically from its screenshot); the subsequent task is to identify the brand of the logo. The latter task can be accomplished by means of DL today, as demonstrated by recent works, e.g., by employing \siamese neural networks~\cite{lin2021phishpedia,phishintention-usenix-sec-2022}. 
Given the relevance of these solutions in anti-phishing schemes, we scrutinize the robustness of DL models for logo identification against subtle adversarial perturbations. Even though many efforts in the DL community reveal the vulnerability of image classification models to adversarial examples~\cite{intriguing,fgsm,poursaeed2018generative,moosavi2017universal}, to the best of our knowledge, there exists no work that studies the vulnerability of logo-based phishing detectors against such sophisticated attacks. 
Therefore, besides the \siamese models proposed by prior work, we also develop two new logo-identification solutions based on state-of-the-art transformer models from Computer Vision---namely, Vision Transformer~\texttt{ViT}~\cite{dosovitskiy2020image} and \texttt{Swin}~\cite{liu2021swin}. 

Subsequently, we propose a novel attack using \textit{generative adversarial perturbations} (GAP)~\cite{poursaeed2018generative}, to craft adversarial logos that simultaneously deceive \lstpar{i}~DL models for logo identification, and \lstpar{ii}~human users, i.e., potential victims. Through a comprehensive experimental study based on datasets of real logos, we demonstrate the quality of our proposed DL models for logo identification and the efficacy of the adversarial logos generated by our GAP attack to evade all three powerful models for logo identification (\siamese, \vit and \swin). 

Finally, we carry out two user studies to assess the impact of our attack on real humans.
We summarise our three major contributions:
\vspace{-0.2cm}

\begin{enumerate}
    \item We propose \textit{a novel attack}, based on generative adversarial perturbations (GAP), against logo-based anti-phishing schemes (Section~\ref{sec:gen-attack}). Our proposed attack treats a phishing detection (specifically, logo-identification) model as a black-box and does not require any model-specific information. 
    \item We propose \textit{two new logo-identification solutions} leveraging transformer-based DL models: \vit and \swin (Section~\ref{sec:logo_models}). We empirically demonstrate that both \vit and \swin achieve performance comparable to the state-of-the-art solutions relying on \siamese models~\cite{lin2021phishpedia,phishintention-usenix-sec-2022} (Section~\ref{sec:eval:base_detection}).

    \item Through a reproducible evaluation on real data, we \textit{evaluate the robustness of three DL models for logo-identification} (\vit, \swin, \siamese) against our GAP-based attack (Section~\ref{sec:eval:adver_attack}). We further validate the \textit{impact of our attack on real humans} through a user study entailing $\sim$250 people (Section~\ref{sec:user}).  
\end{enumerate}
We suggest potential countermeasures against our attack, and also discuss ways that attackers can use to circumvent such countermeasures (Section~\ref{sec:discussions}). Finally, we publicly release our resources to the scientific community~\cite{adversarial-code-2023}.

\vspace{-0.2cm}
\section{Threat model}
\label{sec:threat_model}
\vspace{-0.2cm}
We describe the threat model by first summarizing the functionality of the target system, and then presenting the characteristic of our envisioned attacker.

\vspace{-0.3cm}
\subsection{Target system: Logo-based phishing website detectors}
\label{sec:threat_model_target}
\vspace{-0.25cm}

\begin{figure}[!htbp]
\vspace{-0.3cm}
    \centering
    \includegraphics[width=0.8\textwidth]{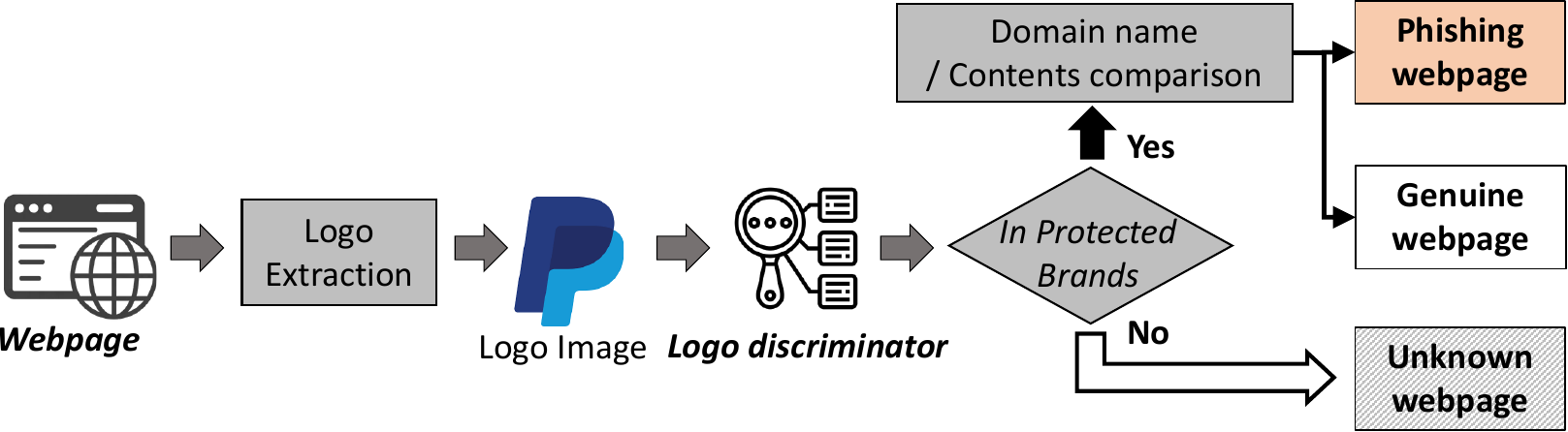}
    \caption{Detection process of logo-based phishing detection systems}
    \label{fig:phishing_detect_arch}
    \vspace{-0.6cm}
\end{figure}

Fig.~\ref{fig:phishing_detect_arch} presents the general workflow of logo-based phishing detection systems. From a given webpage, the detection system first extracts the logo as an image; then, it identifies the brand the logo belongs to by using a discriminator.
Such a discriminator can be implemented in various ways, e.g., earlier works employed methods based on SIFT (scale-invariant feature transformation)~\cite{phishzoo,verilogo}; however, current state-of-the-art methods use DL models~\cite{lin2021phishpedia,phish-siamese-ORB-2022,phishintention-usenix-sec-2022}, and we focus on these. 
Upon identifying the brand of a logo, the system determines if the webpage is legitimate or not by comparing the webpage's domain with the domain of the brand associated with the logo.

Since logo-identification is a multi-class classification problem, the DL model is trained on a static set of classes, i.e., the brands of the logos. Such a set of {\em protected brands} determines the size of the prediction classes; one brand may have multiple logos. 
Previous research has shown that 99\% of the attacks target less than 200 brands~\cite{visualphishnet,lin2021phishpedia,phishintention-usenix-sec-2022}. 

In practice, phishing detectors must exhibit low false-positive rates (FPR), typically below $10^{-3}$~\cite{d-fence-2021,quiring2022phishing}. To successfully detect phishing webpages while maintaining low FPR, logo-based detectors follow two principles~\cite{lin2021phishpedia}: \lstpar{a}~the highest predicted class is decided as the target brand {\em if and only if} the prediction probability is greater than a predefined {\em decision threshold} (say, $\theta$); \lstpar{b}~if the identified logo does not belong to any brand in the protected set, the webpage is considered benign to avoid triggering false positives (see Fig.~\ref{fig:phishing_detect_arch}).
Unfortunately, these principles can be maliciously exploited: by lowering the prediction probability, it is possible to evade logo-based phishing detectors.

\vspace{-0.2cm}
\subsection{Attack: Adversarial logos}
\label{sec:threat_model_attack}
\vspace{-0.2cm}
The basic intuition behind our attack is to create an {\em adversarial logo} that is \lstpar{i}~minimally altered w.r.t. its original variant (to deceive the human eye); and that \lstpar{ii} misleads the phishing detector. Let us describe our attacker by using the well-known notion of adversarial ML attacks~\cite{Biggio:Wild,apruzzese2023position}.

\begin{itemize}
    \item \underline{Goal}: The attacker wants to craft an adversarial logo related to brand $b$ which evades the phishing detector (at inference) while deceiving human eyes.

    \item \underline{Knowledge and Capabilities}:
To train a model for evasion, an attacker can collect authentic logos of any brand (e.g., of PayPal), via crawling or from public datasets (e.g., \textit{Logo2K+}~\cite{wang2020logo}).
The attacker knows that their victims are protected by a logo-based phishing detector powered by ML. 
The attacker has a way to infer the decision result of the phishing detector (this is doable even if the detector is ``invisible''~\cite{apruzzese2023position}, e.g., by inspecting visits to the hosted phishing webpage). The attacker \underline{does not} i)~require knowledge of the logo-identification model employed by the phishing detector, ii)~manipulate the data used to train the ML model. In other words, it's neither a white-box attack nor performs data poisoning. 

Note, the attacker targets a set of brands for phishing; if the targeted brand is not within the protected set, then that is already favorable for an attacker---there is no perturbation required!
Finally, the attacker naturally has control on their phishing webpages.

\item \underline{Strategy}: The attacker manipulates the logo(s) of brand $b$ in their phishing webpages by introducing perturbations so that the logo-identification model predicts with lower confidence, i.e., the probability of the logo being of any brand is lower than the decision threshold ($\theta$). This way, the phishing detector decides the logo {\em not} to be one of the protected brands, which makes way for successful evasion.
\end{itemize}

\noindent \textbf{Scope of attack.}
In our threat model, the attacker exploits the vulnerability of \textit{logo-identification} methods integrated into phishing detectors. 
We focus on logo-identification DL models because they are i)~state-of-the-art research with  phishing detecting capability in the wild (`zero-day' phishing)~\cite{lin2021phishpedia,phishintention-usenix-sec-2022}, and ii)~used in commercial phishing detectors~\cite{apruzzese2023position}.
Threats against logo extraction from a webpage, however interesting, are not within the scope of our current work. 
Lastly, we do not consider attacks to make an unknown logo be identified as one of the protected logos, as that is not beneficial for the attacker.
\section{Deep Learning for Logo-based Phishing Detection}\label{sec:logo_models}

Development of the transformer architecture~\cite{vaswani2017attention} paved the way for various state-of-the-art language models, such as BERT, ChatGPT, and PaLM. Dosovitskiy et al.~\cite{dosovitskiy2020image} applied transformer to Computer Vision tasks with the introduction of Vision Transformer (\vit), demonstrating state-of-the-art performance on benchmark datasets~\cite{dosovitskiy2020image}. The attention mechanism in transformers allows them to capture local and global contextual information effectively, resulting in superior performance on large-scale image classification tasks. This capability is also beneficial for logo identification, since logos of the same brand, while being visually distinct, share the same inherent design structure. Therefore, in this work, we propose, develop and evaluate two transformer-based models, \vit and \swin, for logo identification. To the best of our knowledge, we are the first to leverage transformers for logo-based phishing detection.

\begin{figure}[!htbp]
\begin{minipage}[b]{0.48\linewidth}
\centering
\includegraphics[width=\textwidth]{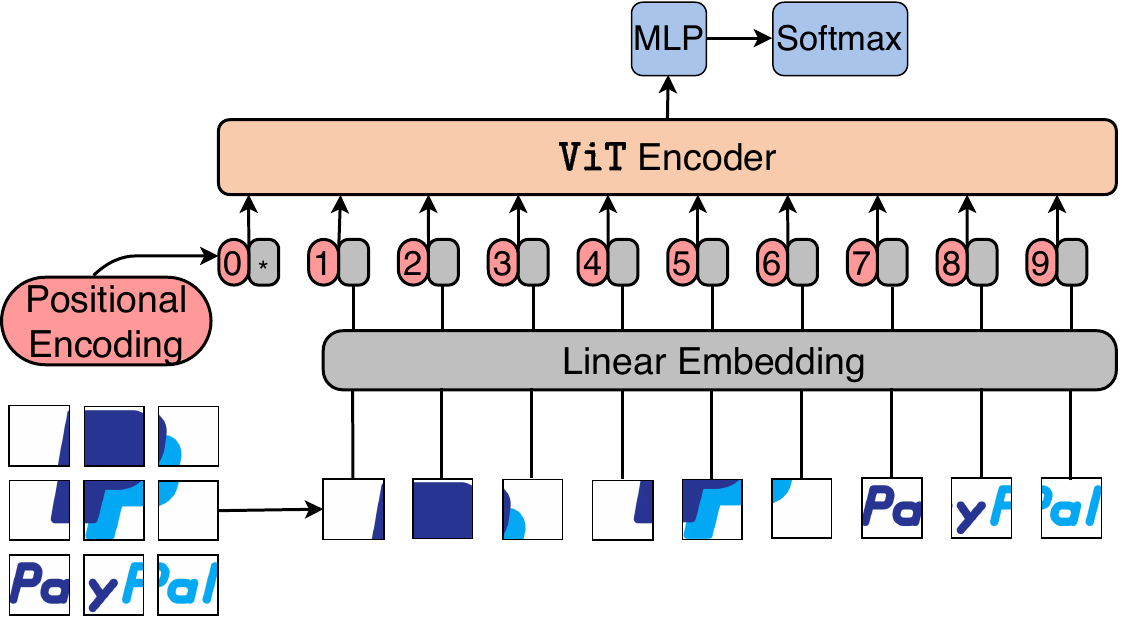}
\caption{\vit-based Model Architecture}
\label{fig:vit_arch}
\end{minipage}
\hspace{0.2cm}
\begin{minipage}[b]{0.48\linewidth}
\centering
\includegraphics[width=\textwidth]{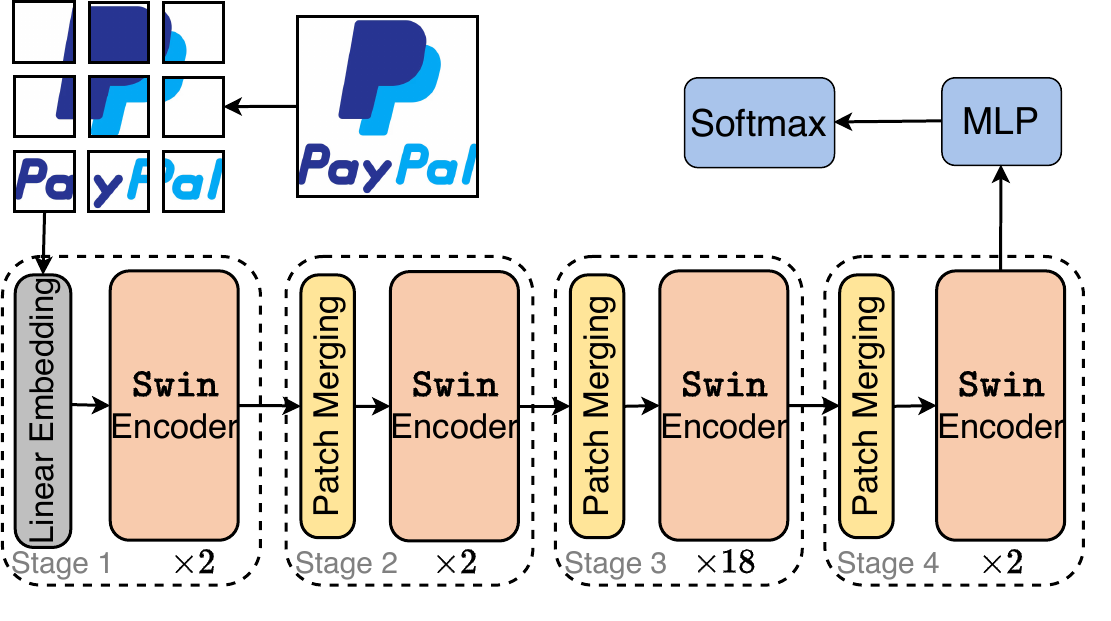}
\caption{\swin-based Model Architecture}
\label{fig:swin_arch}
\end{minipage}
\vspace{-1cm}
\end{figure}

We now describe our proposed \vit (Section~\ref{subsec:vit}) and \swin (Section~\ref{subsec:swin}), for which we provide an overview in Figs~\ref{fig:vit_arch} and~\ref{fig:swin_arch}. Then, we present our own implementation of \siamese (Section~\ref{subsec:siamese}) neural networks. Altogether, these three DL models will represent the target of our attacks (Section~\ref{sec:eval}).

\vspace{-0.3cm}

\subsection{\vit for logo identification}
\label{subsec:vit}

As illustrated in Fig.~\ref{fig:vit_arch}, we develop a logo-identification model by fine-tuning a pre-trained \vit-base model~\cite{dosovitskiy2020image} on our dataset (which we discuss in Section~\ref{sec:eval:dataset}). The model takes as input an image of size $3 \times 224 \times 224$. The image is then split into patches, each of size $16 \times 16$, for further processing. Each patch is then linearly embedded into a vector of size $1 \times 768$. An additional classification token is then added to the linear embedding to form an embedded vector of size $197 \times 768$. The embeddings are positionally encoded before being fed into the transformer encoder. 
Finally, a fully connected layer takes the output from the encoder and maps it to a 2-dimensional space.  The resulting logits are passed through a softmax layer to produce the final prediction probabilities for each class (logo). We denote this new logo-identification model as $\mathcal D_\text{\texttt{ViT}}$.

\vspace{-0.3cm}
\subsection{\swin for logo identification}
\label{subsec:swin}
\vspace{-0.12cm}

Next, we propose \swin-based logo-identification model that utilizes the \swin transformer, a hierarchical transformer architecture introduced by Liu et al.~\cite{liu2021swin}.  Unlike \vit, \swin uses shifted windows to efficiently compute local self-attentions and build hierarchical feature maps through patch merging techniques. As illustrated in Fig.~\ref{fig:swin_arch}, each window contains multiple non-overlapping patches, and each transformer block in the \swin architecture contains two attention layers: a window-based multi-head self-attention (W-MSA) layer that calculates local attention within a specific window, and a shifted window-based multi-head self-attention (SW-MSA) layer that introduces cross-window connections. This approach allows for more efficient computation while still extracting both local and global contextual information.

In our implementation, we use the \texttt{\swin-Transformer-Small} architecture proposed by Liu et al.~\cite{liu2021swin}. The model takes an input image of size $3 \times 224 \times 224$, which is split into patches of size $4 \times 4$. As depicted in Figure~\ref{fig:swin_arch}, the patches are fed sequentially into four encoding stages consisting of 2, 2, 18, and 2 encoder blocks. Each encoding stage merges and downsamples the size of the feature maps by a factor of two, while doubling the number of channels.

The final feature map of size is $7 \times 7$ is transformed by a fully connected and softmax layer to obtain the output logits. We denote this model as $\mathcal D_\text{Swin}$.

\vspace{-0.2cm}
\subsection{\siamese and $\text{\siamese}^{++}$ for logo identification}
\label{subsec:siamese}

The \siamese neural network is a state-of-the-art for image-based phishing detection, both for comparing screenshots~\cite{visualphishnet} and logos~\cite{lin2021phishpedia,phish-siamese-ORB-2022,phishintention-usenix-sec-2022}. 
In logo-based phishing detectors, \siamese models measure the similarity of a given logo to those in the protected set. 
We train a \siamese model as proposed in Phishpedia~\cite{lin2021phishpedia} and PhishIntention~\cite{phishintention-usenix-sec-2022},
utilizing a transfer learning approach.
Specifically, we train a logo classification model with the \texttt{ResNetV2} network as the backbone, which effectively extracts different features from various logo variants. We then connect the trained \texttt{ResNetV2} network to a Global Average Pooling layer to output a vector for any given logo.  The learned vector representation is compared to those of the logos of protected brands using cosine similarity; the target with the highest similarity is identified as the brand the logo is trying to imitate. 

We refer to our implementation of the \siamese model as $\mathcal D_\text{Siamese}$. Additionally,  Phishpedia~\cite{lin2021phishpedia} proposed an adversary-aware detector by replacing the ReLU activation function with a variant called step-ReLU (Appendix~\ref{app:step-relu}). We also consider this robust version of \siamese, which we refer to as $\mathcal D_{\text{Siamese}^{++}}$.

\vspace{-0.25cm}
\section{Our Attack: Adversarial Logos}
\label{sec:gen-attack}
\vspace{-0.1cm}

While recent logo-based phishing detection systems~\cite{lin2021phishpedia,phishintention-usenix-sec-2022} have demonstrated robustness against generic gradient-based attacks such as FGSM~\cite{fgsm} and DeepFool~\cite{moosavi2016deepfool},\footnote{FGSM and DeepFool assume an adversary with complete knowledge of the target classifier, which is much stronger (and less realistic~\cite{apruzzese2023position}) than the attacker envisioned in our threat model.} their resilience against more sophisticated adversarial attacks proposed in the literature~\cite{poursaeed2018generative,moosavi2017universal} remains unexplored. To this end, we propose a DL-based generative framework inspired by Generative Adversarial Perturbations (GAP)~\cite{poursaeed2018generative}, that specifically trains against logo identification models. This framework generates perturbation vectors that can be added to a target logo image, allowing the perturbed logo to evade phishing detection while remaining imperceptible to the human eye.
We now describe our framework at a high-level (Section~\ref{sec:framework}), for which we provide an overview in Fig.~\ref{fig:generator_arch}; and then provide low-level details on how to practically implement our attacks (Section~\ref{sec:implementation}).

\vspace{-0.2cm}
\subsection{Framework: generative adversarial perturbations for logos}\label{sec:framework}
\begin{figure}[t]
    \centering
    \includegraphics[width=0.8\textwidth]{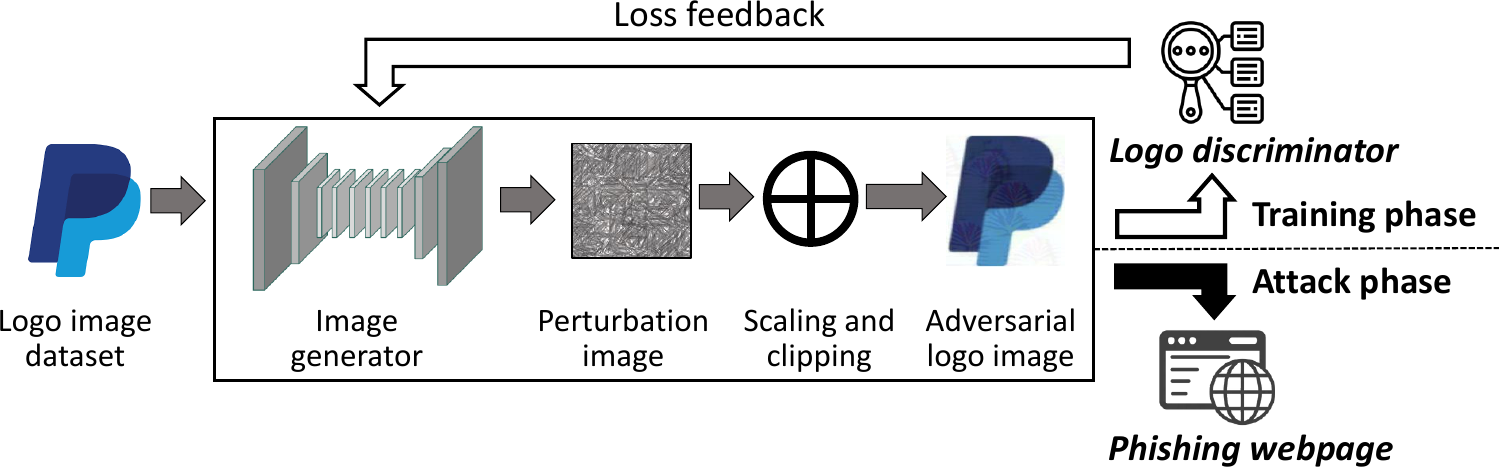}
    \caption{Generative adversarial perturbation workflow}
    \label{fig:generator_arch}
    \vspace{-0.5cm}
\end{figure}

As illustrated in Fig.~\ref{fig:generator_arch}, our framework involves training a \generator that learns to generate perturbations. When added to a logo image, these perturbations can mislead a logo-identification model, which acts as the \discriminator, into lowering its prediction probability below the decision threshold. During the training process, the weights of the \discriminator are frozen, treating it as a black box to guide the training of the \generator.

\textbf{Generator.}
We employ a Deep Residual Network with six residual blocks (ResNet-6)~\cite{he2016deep} as the core architecture of our \generator. Given a legitimate logo image as input, the \generator is trained to generate a {\em perturbation vector}.
The generated perturbations undergo a \textit{Scaling and Clipping} stage. In this stage, the perturbation vector is first scaled and normalized based on the $L_\infty$ norm to control the magnitude of the perturbations, so that they remain imperceptible to human viewers. Subsequently, the normalized perturbations are added pixel-wise to the legitimate logo image, resulting in the adversarial logo.

\textbf{Discriminator.}
The \discriminator is a pre-trained multi-class classifier designed to process a logo image and estimate the probability that the image belongs to a target brand in the protected set. In our framework, we select one of the logo-identification models described in Section~\ref{sec:logo_models} to serve as the \discriminator. 

\vspace{-0.2cm}
\subsection{Implementation}
\label{sec:implementation}
We utilize the pre-trained \discriminator as a black box to assess the effectiveness of the \generator in crafting adversarial logo images. The \discriminator predicts the probability of a given logo belonging to each of the $k$ protected brands; $\mathbf V_\text{true}:[p_1, p_2, p_3....p_k]$, where $\sum_{i=1}^{k} p_{i} = 1$. 
As mentioned in Section~\ref{sec:threat_model_target}, for a webpage to be classified as phishing, the logo-identification model must confidently identify the logo as one of the target brands $i$ from the protected set, with a probability $p_i$ greater than the phishing detector's decision threshold $\theta$. 

Hence, to devise our \generator, we introduce a target probability $p_\text{adversarial}$, such that $p_\text{adversarial} < \theta$. The \generator is trained to craft adversarial logos that are classified with probabilities lower than $p_\text{adversarial}$ for all of the protected brands, so as to evade phishing detection.
Empirically, we observe that $\theta$ is very high (above 0.8) for all discriminators, and for our attacks, $p_\text{adversarial}$ can be much lower (in our experiments, it is 0.5; see Table~\ref{tab:eval:gen_training} in Appendix~\ref{appendix:config}). 

To guide the training process, the \generator is trained with a target probability vector $\mathbf V_\text{target}:[p_1', p_2', p_3'....p_k']$, where each element $p_i'$ is defined such that $p_i' = \min(p_i, p_\text{adversarial})$.  This ensures that the generated adversarial logos are classified with probabilities below the $\theta$ for all protected brands.

The loss function is defined as a decreasing function of the cross entropy $\mathcal H(V_\text{true}, V_\text{target})$ between the target probability vector $\mathbf V_\text{target}$ and $\mathbf V_\text{true}$. The specific form of the loss function can be expressed as follows:
\begin{equation}
    \text{loss} = \log\left(\mathcal H\left(\mathbf V_\text{true}, \mathbf V_\text{target}\right)\right)  
\end{equation}

Minimizing this loss, the \generator learns to craft adversarial logos that evade phishing detection\footnote{\textbf{Remark:} Our attack relies on the logos generated by the \generator, which in turn depend on a \discriminator, i.e., a DL model for identifying logos. However, the \discriminator \textit{does not} necessarily have to be the identical one used in the targeted phishing detection system: as our experiments show, our adversarial logos evade even DL models that have not been used to develop the \generator (by leveraging the well-known transferability property of adversarial examples~\cite{demontis2019adversarial}).}; furthermore, perturbations preserve the visual similarity with the original logo, thereby facilitating deception to the human eye.

\vspace{-0.25cm}
\section{Experimental evaluations}\label{sec:eval}

We now empirically assess the quality of our contributions.  We begin by describing the datasets used for our experiments (Section~\ref{sec:eval:dataset}), and introduce the metrics used for our performance assessment (Section~\ref{sec:eval:metrics}). Then, we first show that our two DL models for logo-identification achieve state-of-the-art performance (Section~\ref{sec:eval:base_detection}), and then demonstrate that our attacks can evade all our considered logo-identification models (Section~\ref{sec:eval:adver_attack}).  Our code, dataset used, as well as generated perturbed logos are available at~\cite{adversarial-code-2023}.

\vspace{-0.25cm}
\subsection{Dataset}\label{sec:eval:dataset}
To evaluate the performance of logo-based phishing detectors and their robustness against generative adversarial perturbations, 
we use two sets of logo images:

\begin{itemize}
    \item $\mathbf{L}$, \textbf{Protected brands:} The logo image set of protected brands, $\mathbf{L}$, consists of images of $181$ brands which are identical to the brands used in Phishpedia~\cite{lin2021phishpedia}. According to the empirical observation in~\cite{lin2021phishpedia}, 
    $99\%$ of phishing pages target one of these $181$ brands.
    For these protected brands, we collected $28\,263$ public logo images from search engines and Pawar's logo image dataset~\cite{revanks2021logo}. Each brand's logo has $100$--$200$ variants.
    \item \textbf{$\mathbf{\bar{L}}$, Unprotected brands:} Logo image set $\mathbf{\bar{L}}$ is the set of $2\,045$ images from $2\,000$ brands that do not belong to the brands in $\mathbf{L}$. The image samples are from the \textit{Logo2K+} dataset, which is publicly available~\cite{wang2020logo}.
\end{itemize}
The data was collected in the second half of January 2023.

\vspace{-0.2cm}
\subsection{Performance Metrics}\label{sec:eval:metrics}
In what follows, we denote the logo-identification models as  \textbf{\texttt{discriminators}}; the attack \texttt{generators} also use the discriminators in their training phase.

\text{\underline{Logo identification performance:}}
We provide the definitions of metrics for logo-based phishing webpage detection.  Note that, for a discriminator used for phishing detection, the positives are the logos in $\mathbf{L}$, the protected brand list, that need to be identified. 
If the highest prediction probability of a logo is below a certain decision threshold, it is classified as an unknown brand.

\begin{itemize}
\item{\textit{True positive (TP)}:} A TP in our evaluation denotes the case of correct brand identification of the given logo (of a protected brand)  by the discriminator.  
\item{\textit{False positive (FP)}:} An FP denotes the case when the given logo image is wrongly identified as one of the protected brands when in reality, the given logo image does not belong to the protected brand set.  
\item{\textit{True negative (TN)}:} A TN occurs when the brand of the given logo is not in the protected brand set and gets correctly classified as an unknown brand.
\item {\textit{False negative (FN)}:} An FN denotes when the brand of the given logo belonging to the protected brand set is classified as any other brand.

\end{itemize}

\begin{comment}
Denoting the actual brand of a given logo $l$ as $l_{b}$, the predicted brand by the discriminator as $l_{p}$, the above metrics are formalized as: 
\begin{equation}
    \text{eval\_verdict}(l_{p}) =     
        \begin{cases}
        \text{TP}, & \text{for } l_{p} = l_{b} \land l_{b} \in \mathbf{L}  \\        
        \text{FP}, & \text{for } l_{p} \in \mathbf{L} \land l_{b} \in \mathbf{\bar{L}} \\
        \text{TN}, & \text{for } l_{p} \notin \mathbf{L} \land l_{b} \in \mathbf{\bar{L}} \\
        \text{FN}, & \text{for } l_{p} \neq l_{p} \land l_{b} \in \mathbf{L}
        \end{cases}
\end{equation}
\noindent
We can then 
\end{comment}
\noindent
Denoting the actual brand of a given logo $l$ as $l_{b}$, and 
the predicted brand by the discriminator as $l_{p}$,
we define the True Positive Rate (TPR) and False Positive Rate (FPR):

\vspace{-0.5cm}
\noindent\begin{tabularx}{0.9\textwidth}{@{}XXX@{}}
\begin{equation*}
    \text{TPR} = {|(l_{b} = l_{p}) \land (l_{p} \in \mathbf{L})| \over |l_{b} \in \mathbf{L}|}{;}\label{eq:tpr}
\end{equation*} &
\begin{equation}
    \text{FPR} = {|(l_{p} \in \mathbf{L}) \land (l_{b} \in \mathbf{\bar{L}})| \over |l_{b} \in \mathbf{\bar{L}} |}\label{eq:fpr}
    \end{equation} \\
\end{tabularx}

\text{\underline{Impact of the attacks:}} Recall that our attacker aims to fool the discriminator into classifying a protected brand logo as an unknown brand. Hence, we introduce the \textit{Fooling ratio}, which is the rate of adversarial logos classified as being of an unknown brand (out of all the phishing logos). Formally:
\begin{equation}
\text{Fooling ratio} = {|l_{p} \notin \mathbf{L} \land l_{b} \in \mathbf{L}| \over |l_{b} \in \mathbf{L} |}\label{eq:fooling}
\end{equation}
Intuitively, a higher fooling ratio denotes an attack with a higher impact.

\subsection{Baseline: Analysis of logo-identification models}
\label{sec:eval:base_detection}
We assess the performance of the four DL models for logo-identification presented in Section~\ref{sec:logo_models}. Specifically, we first measure the TPR and FPR of the state-of-the-art discriminators (i.e., \texttt{Siamese} and its robust version \texttt{Siamese}$^{++}$~\cite{lin2021phishpedia}), and compare them with the transformer-based discriminators that we proposed in this work (i.e., \texttt{ViT} and \texttt{Swin}).

\underline{Setup.} We use the datasets $\mathbf{L}$ and $\mathbf{\bar{L}}$ (see Section~\ref{sec:eval:dataset}), with a train:test split of 85:15. For \texttt{ViT} and \texttt{Swin}, we apply the common model head fine-tuning for 50 epochs and then transfer training on the entire networks for the next 150 epochs, reducing computational time while improving performance.
We provide hyperparameters configurations of our discriminators in Table~\ref{tab:eval:disc_training} (in the appendix).

\underline{Results.} Fig.~\ref{fig:logo_identification_tpr_fpr} shows the ROC curves of the four discriminators (the x-axis denoting FPR is in log-scale for visibility).  Overall, \texttt{Siamese} and \texttt{Siamese}$^{++}$ show the best performance in terms of logo identification.  All four models show comparable TPRs at FPR above $10^{-2}$. For practical purposes, however, we have to evaluate the detection capability at low FPRs~\cite{quiring2022phishing,divakaran2022phishing}. Observe that, the TPR values of the discriminators \texttt{ViT} and \texttt{Swin} at FPR below $10^{-2}$ are worse than the \siamese models. Fig.~\ref{fig:logo_identification_tpr_fixed_fpr} shows the gap in TPR between the discriminators at the more practical FPR value of $10^{-3}$; \texttt{Siamese} and \texttt{Siamese}$^{++}$ show around six and twelve percent-point higher TPR than the \texttt{ViT} and \texttt{Swin}, respectively. 

\vspace{-0.8cm}
\begin{figure}[!htbp]
    \centering
    \subfloat[ROC curves]{\label{fig:logo_identification_tpr_fpr}
        \includegraphics[width=0.385\textwidth]{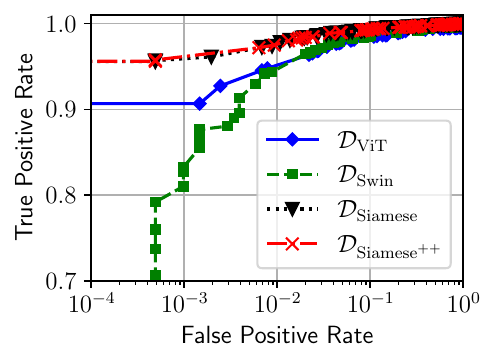}}
    \subfloat[TPR at $10^{-3}$ FPR]{\label{fig:logo_identification_tpr_fixed_fpr}
        \includegraphics[width=0.38\textwidth]{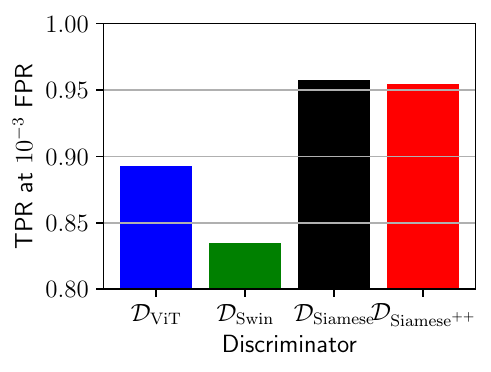}}
    \caption{Comparing discriminators for logo identification}    
    \vspace{-0.65cm}
\end{figure}

Although \swin and \vit are not better than \siamese, they still achieve an appreciable degree of performance, and hence are used to evaluate our attacks.

\vspace{-0.2cm}
\subsection{Attack: evasiveness of adversarial logos, and computational cost}\label{sec:eval:adver_attack}
\vspace{-0.1cm}

We quantitatively analyze the effects of adversarial logos generated by our attack against DL models for logo identification. We do this through a cross-evaluation that captures both `white-box' and `black-box' adversarial settings. At the end of this section, we also discuss the computational cost of our attacks.

\underline{Setup.} Recall that our attack (Section~\ref{sec:gen-attack}) entails training a generator by using a given discriminator (i.e., DL models for identifying logos). For our experiments, we consider three discriminators: \texttt{ViT}, \texttt{Swin} and \texttt{Siamese}, thereby yielding three corresponding generators: $\mathcal G_{\text{ViT}}$, $\mathcal G_{\text{Swin}}$ and $\mathcal G_{\text{Siamese}}$. After training each generator, we assess the adversarial logos against \textit{all our discriminators}. 
Such an evaluation protocol allows one to analyze the effects of our attacks when the adversary does not know the DL model used for the defense. 

For evaluations, we train our generators on the dataset $\mathbf{L}$; 
we provide the hyperparameters of our generators in Table~\ref{tab:eval:gen_training} (Appendix~\ref{appendix:config}). Subsequently, we test the discriminators with the adversarial logos crafted by each generator.

\begin{figure}[!htbp]
\vspace{-0.8cm}
    \centering
    \subfloat[$\mathcal{G}_\text{ViT}$]{\label{fig:cross_compare_ViT}
        \includegraphics[width=0.26\textwidth]{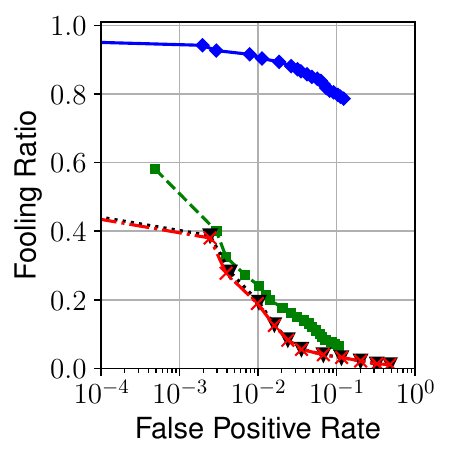}}
    \subfloat[$\mathcal{G}_\text{Swin}$]{\label{fig:cross_compare_Swin}
        \includegraphics[width=0.24\textwidth]{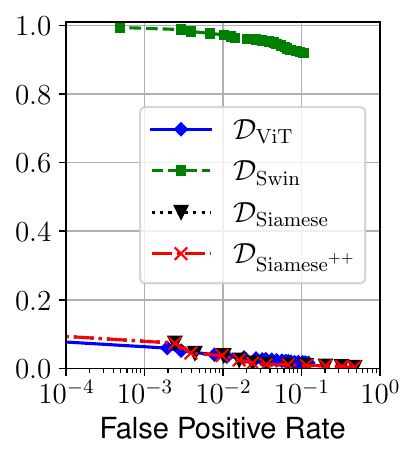}}
    \subfloat[$\mathcal{G}_\text{Siamese}$]{\label{fig:cross_compare_Siamese}
        \includegraphics[width=0.24\textwidth]{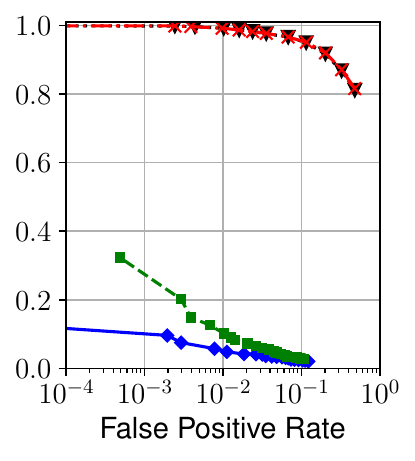}
    }
    \subfloat[at $10^{-3}$ FPR]{\label{fig:cross_compare_bar}
        \includegraphics[width=0.228\textwidth]{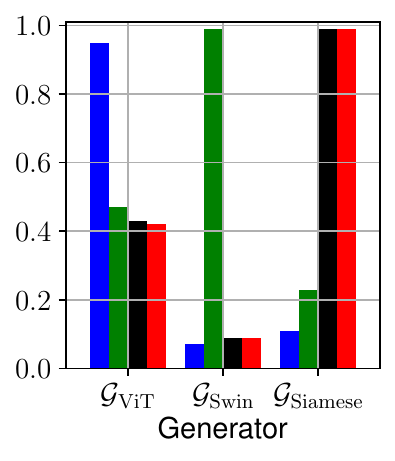}
    }
    \caption{Comparison of different generators against different discriminators}
    \label{fig:cross_compare}
 \vspace{-0.5cm}   
\end{figure}

\underline{Results.} The results are plotted in Fig.~\ref{fig:cross_compare}, where we compare the fooling ratio of discriminators against the different attacker models for varying FPRs (in log-scale). It stands out that each discriminator is much weaker against the adversarial logos created by the `matching' generator compared to those created by generators trained on different discriminators.  For instance, from Fig.~\ref{fig:cross_compare_ViT}, we observe that the adversarial logos generated by $\mathcal{G}_{\text{ViT}}$ are more effective against \vit (blue line) than against \swin (green line).
We observe from Fig.~\ref{fig:cross_compare_Swin} and Fig.~\ref{fig:cross_compare_Siamese} that, if the attacker's generator model is not trained with \texttt{ViT}, the fooling ratio drops significantly for the defender with the \texttt{ViT} discriminator.

From the adversary's perspective, \texttt{ViT} is the most effective generator against all discriminators.
Fig.~\ref{fig:cross_compare_bar} compares the fooling ratios of the four discriminators at a fixed FPR of $10^{-3}$; note, \textbf{fooling ratios against $\mathcal G_\text{ViT}$ are high, ranging from 42\% to~95\%}. In other words, with $\mathcal G_\text{ViT}$, at least 42\% of attacker generated logos can evade phishing detectors, independent of the discriminator deployed.
Against such an attacker, the defender might prefer to use \texttt{Siamese} (or \texttt{Siamese}$^{++}$) as it achieves the lowest fooling ratio (of around 42\% at $10^{-3}$ FPR). 
Interestingly, the most robust model for the defender against an {\em arbitrary} generator model would be \texttt{ViT}, since, on average, \texttt{ViT} achieves a lower fooling ratio against all generator models. 

\underline{Computational cost.} Two factors contribute to the computation time to realize our adversarial logos: i)~ generator training and ii)~perturbed logo generation. We measure the generator training time with the three models, i.e., \texttt{ViT}, \texttt{Swin}, and \texttt{Siamese}, for each training epoch and the required epochs till reaching a compelling performance, i.e., 0.9 of fooling ratio against the discriminator with the corresponding model. The experiments are performed on a system with NVIDIA RTX3090 GPU, 2.8GHz 32-core AMD CPU, 80GB RAM with Python 3.8.10, and PyTorch 1.2.0 on Ubuntu~20.04~OS. We report the results in Table~\ref{tab:generator_training_time}.

From this table, we observe an apparent gap between the models in their training time. While the \texttt{ViT}-based generator, $\mathcal{G}_\text{ViT}$, takes only half the training time per epoch in comparison to $\mathcal{G}_\text{Swin}$, it requires five times more training epochs to reach the same level of performance, (i.e., 0.9 fooling ratio).  $\mathcal{G}_\text{Siamese}$ shows significantly less overhead than the other two, in both, training time per epoch and the required epoch.  $\mathcal{G}_\text{Siamese}$ accomplishes a fooling ratio of 0.9 against $\mathcal{D}_\text{Siamese}$ after just one epoch of training which takes only eight minutes.  Overall, training $\mathcal{G}_\text{ViT}$ takes 744 minutes to have 0.9 fooling ratio, which is around 2.8 and 93 times longer training time than $\mathcal{G}_\text{Swin}$ and $\mathcal{G}_\text{Siamese}$, respectively. Although there are significant differences in training times, when it comes to generating perturbed logos, all three generators take only around 0.7 seconds per image on average; this negligible cost allows an attacker to generate a large number of samples to test against a deployed phishing detector.

\begin{table}[t]
    \caption{Training time for the perturbation generators}
    \centering    
    \begin{tabular}{@{\hspace{1em}} l @{\hspace{1em}} | @{\hspace{1em}}  c @{\hspace{1em}} c @{\hspace{1em}} c @{\hspace{1em}}}    
    \toprule
      & $\mathcal{G}_\text{ViT}$ & $\mathcal{G}_\text{Swin}$& $\mathcal{G}_\text{Siamese}$\\
    \midrule
    Avg. training time per epoch (min.)         & 12    & 23   & 8  \\
    No. of epochs for 0.9 fooling ratio         & 62    & 12   & 1  \\    
    Training time for 0.9 fooling ratio (min.)  & 744   & 277  & 8  \\
    \bottomrule
    \end{tabular}
    \label{tab:generator_training_time}
    \vspace{-0.3cm}
\end{table}

\begin{cooltextbox}
\textsc{\textbf{Takeaways.}}
i)~An attacker with knowledge of the discriminator used for defense achieves more than 95\% fooling ratio with our adversarial generator. ii)~In the absence of knowledge of the discriminator (i.e., independent of the discriminator), an attacker choosing $\mathcal G_{\text{ViT}}$ as the generator achieves a fooling ratio of at least 42\% against the defender (see Fig.~\ref{fig:cross_compare_bar}).
\end{cooltextbox}
\vspace{-0.2cm}
\section{User study: do adversarial logos trick humans?}
\label{sec:user}
\vspace{-0.2cm}
We now provide a complementary evaluation of our proposed attack. Specifically, we seek to investigate \textit{if our adversarial logos can be spotted by humans}. Indeed, even if a phishing detector can be evaded, this would be useless if the human, the actual target of the phishing attack, can clearly see that something is ``phishy''.
Hence, we carry out \textbf{two user-studies}, which we describe (Section~\ref{subsec:user-method}) and discuss (Section~\ref{subsec:user-results}) in the remainder of this section.

\vspace{-0.2cm}
\subsection{Methodology}
\label{subsec:user-method}
\vspace{-0.12cm}
Our goal is to assess if the perturbations entailed in an adversarial logo can be recognized by humans. There are many ways to perform such an assessment through a user-study, each with its own pros and cons\footnote{Designing bias-free user-studies in the phishing context is an open problem~\cite{sharma2021impact,alsharnouby2015phishing}.}.

We build our user-studies around a central research question (RQ): \textit{given a pair of logos (i.e., an `original' one, and an `adversarial' one), can the human spot any difference?} 
Our idea is to design a questionnaire containing multiple pairs of logos, and ask the participants to rate (through a 1--5 Linkert scale) the similarity of the logos in each pair. Intuitively, if the results reveal that users perceive the logos to be ``different'', then it would mean that our adversarial logos are not effective against humans. 

To account for the fact that the responses we would receive are entirely subjective, we carry out (in April 2023) two quantitative user studies:
\begin{enumerate}
    \item \textit{Vertical Study} (VS), which entails a small population (N=30) of similar users (students of a large university, aged 20--30). The questionnaire has ten questions (each being a pair of logos to rate), wherein each participant is shown a different set of questions. The purpose of VS is to capture the responses of a specific group of humans across a large set of adversarial logos.
    \item \textit{Horizontal Study} (HS), which entails a large population (N=287) of users with diverse backgrounds (Amazon Turk Workers with 95+\% hit-rate, aged 18--70). The questionnaire includes 21 questions, which are always the same for each participant. The purpose of HS is to capture the response of various humans to a small set of adversarial logos.
\end{enumerate}
For both VS and HS, participants were asked to provide a response within 5s of seeing the pair of logos (because, realistically, users do not spend much time looking at the logo on a website). We also included control questions (e.g., pairs of identical logos, and pairs of clearly different logos) as a form of attention mechanism\footnote{For HS, we received 322 responses, but we removed 35 because some users took too little time to answer the entire questionnaire, or did not pass our attention checks.}.
Finally, we shuffled the questions to further reduce bias. For transparency, we provide our questionnaire at~\cite{adversarial-code-2023}.

For VS (resp. HS), we included 2 (resp. 3) ``identical'' pairs as baseline; and~5 (resp. 12) ``original-adversarial'' pairs to answer our RQ. 

\subsection{Results}
\label{subsec:user-results}

\begin{figure}[!htbp]
    \centering{
    \subfloat[Vertical Study]{    
        \includegraphics[width=0.44\textwidth]{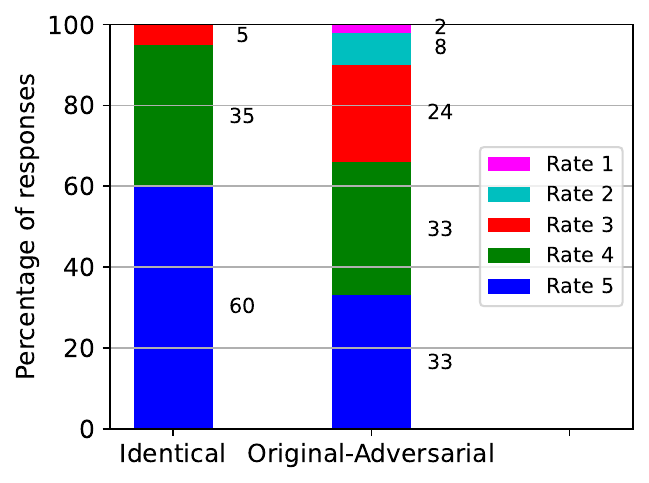}
        \label{sfig:vs}}
    \hfill            
    \subfloat[Horizontal Study]{ 
        \includegraphics[width=0.51\textwidth]{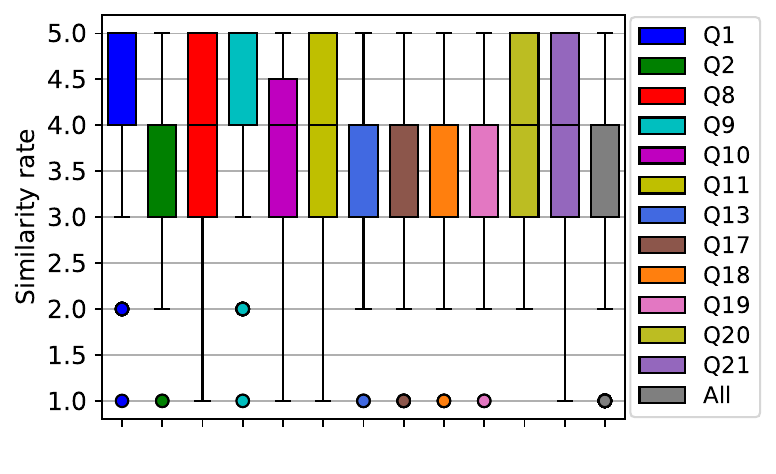}
        \label{sfig:hs}}    
    \caption{Results of our two user-studies: vertical study and horizontal study}
    \label{fig:user-study}
    }
    \vspace{-0.3cm}
\end{figure}

 We present the results of both of our user studies in Fig.~\ref{fig:user-study}. Specifically, Fig.~\ref{sfig:vs} shows the cumulative distribution of the scores for the three `identical' pairs, and the five `original-adversarial' pairs in VS. Whereas the boxplots in Fig.~\ref{sfig:hs} show how the participants of HS rated the 12 ``original-adversarial'' pairs; the rightmost boxplot aggregates all results. In our rating definition, 5 means `similar', and 1 means `different'.

From Figure~\ref{sfig:vs}, we observe that 95\% of all responses (30 users $\times$ 10 questions) rated all `identical' pairs (left bin) between 4 and 5 (only 5\% answered with a 3). That is to say; they correctly guessed that all identical pairs were indeed very similar, thereby also confirming that this population was very reliable. For this reason, we find it noteworthy that \textbf{our adversarial logos are able to deceive them}: in the right bin, 66\% rated the `original-adversarial' pairs with either 4 or 5, and only 10\% rated them with a 1 or 2.

Figure~\ref{sfig:hs} shows the results for the `adversarial-original' pairs (we already removed some clearly noisy answers, as stated in Section~\ref{subsec:user-method}). We observe that the wide majority of HS population rated the pairs as similar (the average is always below the middle point, 3). Hence, we can conclude: HS also reveals that \textbf{our adversarial logos are barely detected by humans as perturbed}.
\vspace{-0.2cm}
\section{Countermeasures (and counter-countermeasures)}
\label{sec:discussions}
\vspace{-0.2cm} 
Given that our adversarial logos can simultaneously fool state-of-the-art DL models for logo-identification and human eyes, we ask ourselves: {\em how can adversarial logos be countered?}
One potential mitigation is to leverage \textit{adversarial learning} by injecting evasive logos in the training set~\cite{apruzzese2020deep}, thereby realizing an {\em adversarially robust} discriminator. However, an expert attacker may anticipate this and can hence attempt to circumvent such a robust discriminator by developing a new generator, thereby crafting more evasive adversarial logos (e.g., as demonstrated in other domains~\cite{shenoi2023ipet,rahman2020mockingbird}). 
We now investigate both of these scenarios through additional proof-of-concept experiments, which involve the strongest discriminator of our evaluation: \vit.

\begin{figure}[!htbp]
    \centering
    \subfloat[Against original generator $\mathcal{G}_{\text{ViT}}$]{
        \label{eval:fig:fooling_org_gen}
        \includegraphics[width=0.45\textwidth]{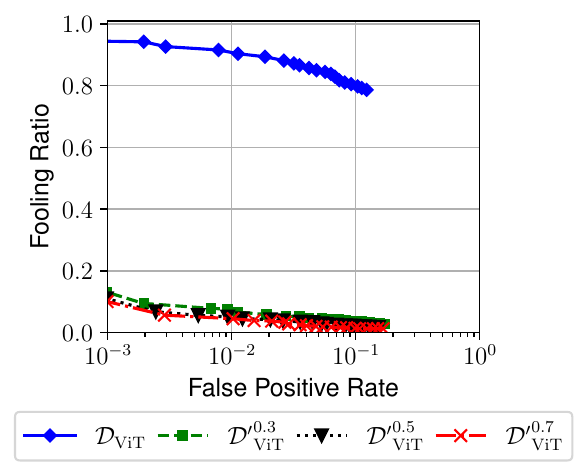}
    }
    \subfloat[Against adaptive generators]{
        \label{eval:fig:fooling_adv_gen}
        \includegraphics[width=0.45\textwidth]{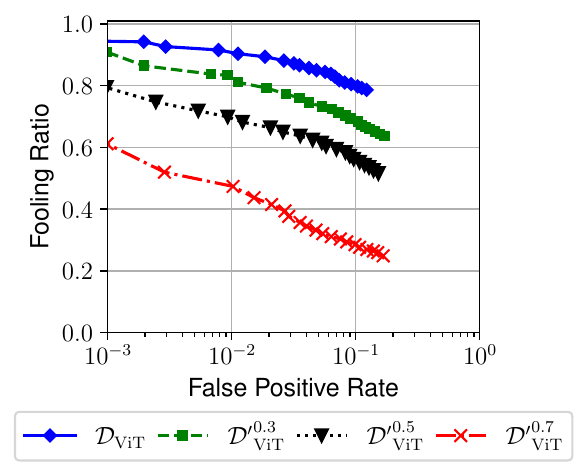}
    }    
    \label{eval:fig:fooling_gen}
    \caption{Performance of discriminator and generator due to adversarial training}
    \vspace{-0.5cm}
\end{figure}

\vspace{0.1cm}
\underline{Countermeasure: building robust discriminator.}
Adversarial training is one of the most well-known techniques to defend against adversarial examples~\cite{shafahi2019adversarial,apruzzese2020deep}. The idea is to update a given ML model by training it on adversarial examples that can mislead its predictions. 
We build our robust discriminators, $\mathcal{D'}^{0.3}_{\text{ViT}}$, $\mathcal{D'}^{0.5}_{\text{ViT}}$, and $\mathcal{D'}^{0.7}_{\text{ViT}}$, by replacing 30\%, 50\%, and 70\% of the logos in the training dataset $\mathbf L$ with their adversarial variants, respectively. In particular, we use the adversarial logos generated with $\mathcal{G}_{\text{ViT}}$, i.e., trained with the vanilla \texttt{ViT} discriminator. Then, we compare these three robust discriminators with the vanilla \texttt{ViT} discriminator $\mathcal{D}_{\text{ViT}}$, against the same attack presented in Section~\ref{sec:eval:adver_attack}. The results are shown in Fig.~\ref{eval:fig:fooling_org_gen}. 
We observe that the robust discriminators exhibit much lower fooling ratios: while the vanilla \texttt{ViT} has a fooling ratio above $0.8$, the robust discriminators have fooling ratios below $0.2$ even at a low FPR of~$10^{-3}$.

\vspace{0.1cm}
\underline{Counter-countermeasure: evading robust discriminators.}
An attacker is also capable of taking a sophisticated strategy to counter a robust logo-identification discriminator built via adversarial training. 
To do this, the attacker must obtain such a robust discriminator---this can be done through well-known black-box strategies~\cite{papernot2017practical,bhagoji2018practical}, or the attacker could even build one on their own. 
The attacker must then use the robust discriminator to train an `adaptive' generator that can yield more evasive perturbations.
For this experiment, we consider the case wherein the attacker trains the adaptive generator by using  $\mathcal{D'}^{0.3}_{\text{ViT}}$, $\mathcal{D'}^{0.5}_{\text{ViT}}$, and $\mathcal{D'}^{0.7}_{\text{ViT}}$, thereby realizing  $\mathcal{G'}^{0.3}_{\text{ViT}}$, $\mathcal{G'}^{0.5}_{\text{ViT}}$, and $\mathcal{G'}^{0.7}_{\text{ViT}}$, respectively.
The results are shown in Fig.~\ref{eval:fig:fooling_adv_gen},  which plots the fooling ratio of the {\em adaptive} generator against the corresponding {\em robust} discriminator.

Compared to the attacks from the `vanilla' generator $\mathcal{G}_{ViT}$ in Fig.~\ref{eval:fig:fooling_org_gen} (which achieves below 20\% of fooling ratio at $10^{-3}$ FPR), the adaptive generators in Fig.~\ref{eval:fig:fooling_adv_gen} are much more effective. Yet, we observe that discriminators trained with more adversarial logos tend to be more robust: at $10^{-3}$ FPR, $\mathcal{D'}^{0.3}_{\text{ViT}}$ has a fooling ratio of 0.9, whereas $\mathcal{D'}^{0.5}_{\text{ViT}}$ and $\mathcal{D'}^{0.7}_{\text{ViT}}$ have 0.8 and 0.6, respectively.

\vspace{-0.8cm}

We find it enticing that this continuous game between attacker and defender, reflected in the generator (attacker) and discriminator (defender), eventually forms the concept of the Generative Adversarial Network (GAN). Indeed, a question rises: ``what happens if this process is repeated many times?'' We plan to address this intriguing research question in our future work. 

\vspace{-0.32cm}
\section{Related works}
\label{sec:background}
\vspace{-0.22cm}

\hspace{0.4cm}
\underline{Phishing Website Detection via ML.}
Many works leveraged statistical models, including ML, for phishing website detection (e.g., \cite{ML-phishing-APWG-2007,Google-content-phishing-2010,canita+-2011,URL-based-ICML-2009,tian2018needle}). Typically, these models are trained on labeled datasets to learn to discriminate between phishing and benign webpages. 
There also exists an orthogonal family of countermeasures, referred to as reference-based phishing detectors, that identify visually similar webpages. This is based on the notion that phishing webpages are more successful when they imitate a legitimate website. This characteristic has been extensively scrutinized by prior literature~\cite{emd-phishing-2006,phishzoo,verilogo,logosense,visualphishnet,logomotive-2022,lin2021phishpedia,phishintention-usenix-sec-2022}. For example, VisualPhishNet trains a \siamese model to detect visually similar \textit{screenshots} between a given webpage and those in a set of well-known brands~\cite{visualphishnet}. Other works (e.g.,~\cite{phishzoo,verilogo,logosense,lin2021phishpedia,phishintention-usenix-sec-2022}) focus on identifying visually invariant \textit{logos}.

\underline{Attacks against ML-based Phishing Website Detectors.}
Expert attackers are aware of the development of anti-phishing solutions and constantly refine their techniques to avoid being taken down. For instance, phishers can use cloaking to evade automated crawlers often used by security vendors~\cite{zhang2021crawlphish}; alternatively, they can exploit `squatting' to evade detectors analyzing the URL~\cite{tian2018needle}. It is also easy to change the HTML contents to evade HTML-based phishing detectors~\cite{phishing-NDSS-MADWeb-2020,apruzzese2022spacephish}. 
Researchers have also examined the impact of adversarial perturbations on image-based phishing detectors~\cite{visualphishnet,lin2021phishpedia,phishintention-usenix-sec-2022,corona2017deltaphish}. However, these attacks assume that the attacker possesses complete knowledge of the deployed model and can access the model gradients, enabling manipulations in the feature-space (for further details, refer to~\cite{apruzzese2022spacephish}). 
We demonstrate a successful attack conducted by an attacker lacking both knowledge of and access to the deployed model. Furthermore, none of the prior works have conducted user studies to validate the practicality of their attacks.

\underline{Adversarial Perturbations.} Moving away from gradient-based perturbations, Moosavi et al. introduced Universal Adversarial Perturbations~\cite{moosavi2017universal}, a framework for learning perturbations that are image-agnostic and generalized across various image classification models. This work sparked further proposals~\cite{shafahi2020universal,mopuri2018generalizable,zhang2020cd} aiming to enhance universal perturbations. Subsequently, Poursaeed et al. proposed Generative Adversarial Perturbations~\cite{poursaeed2018generative}. The generative model achieved state-of-the-art performance, unifying the framework for image-agnostic and image-dependent perturbations and considering both targeted and non-targeted attacks. We draw inspiration from their framework to develop a generative network specifically for crafting adversarial logos.\\

\vspace{0.5mm}
\textbox{
{\small \textbf{Summary.} While prior works have investigated gradient-based attacks~\cite{lin2021phishpedia,phishintention-usenix-sec-2022} against image classifiers, to the best of our knowledge, we are the first to show the feasibility of attacks using a generative neural network model trained to craft adversarial logos, and comprehensively evaluate the impact of such attacks on state-of-the-art methods for logo-identification.}
}

\vspace{-0.2cm}
\section{Conclusions}
\vspace{-0.2cm}
Logo-based phishing detectors have shown significant capabilities with the employment of DL models. In this work, we developed and presented a novel attack against logo-based phishing detection systems. Our experiments demonstrate the capability of an attacker equipped with a generative adversarial model in defeating the detection systems as well as human users. We hope this will trigger further research and development of phishing detection solutions that are robust to adversarial ML attacks. 

\textbf{\textsf{Ethical Statement.}} Our institutions do not require any formal IRB approval to carry out the research discussed herein. We always followed the guidelines of the Menlo report~\cite{bailey2012menlo}. For our user-studies, we never asked for sensitive data or PII. Finally, although we publicly release our code for the sake of science, as mentioned on the GitHub page~\cite{adversarial-code-2023}, such code should not be used for any unethical or illegal purposes.

\textsf{\textbf{Acknowledgment.}} We thank the Hilti Corporation, Trustwave, NUS (National University of Singapore) and Acronis, for supporting this research.

\bibliographystyle{splncs04}
\bibliography{references}

\vspace{-0.5cm}
\appendix
\addcontentsline{toc}{section}{Appendices}
\renewcommand{\thesubsection}{\Alph{subsection}}
\section*{Appendix}

\subsection{Step-ReLu activation Function}
\label{app:step-relu}
The step-ReLU function utilised in  training the robust \siamese model $\mathcal D_{\text{Siamese}^{++}}$ (Section~\ref{subsec:siamese}) is expressed as:

\begin{equation}
    f(x) = \max(0, \alpha \cdot \lceil \frac{x}{\alpha} \rceil)
    \label{step}
\end{equation}

\subsection{Discriminator and generator configurations}\label{appendix:config}

\begin{table}[h]
\vspace{-0.75cm}
\centering
    \caption{Hyperparameter configurations for discriminators}\label{tab:eval:disc_training}  
    \begin{tabular}{@{\hspace{1em}} l @{\hspace{1em}}|@{\hspace{1em}}c@{\hspace{1em}}c@{\hspace{1em}}c@{\hspace{1em}}}
    \toprule
     Parameters & $\mathcal{D}_\text{ViT}$ & $\mathcal{D}_\text{Swin}$& $\mathcal{D}_\text{Siamese}$ \\
    \midrule
    Backbone & ViT & Swin & ResNetV2\\
    Pre-trained Model & ViT-b/16 & Swin-S & BiT-M-R50x1\\
    No. of params & 85.9M & 49.0M & 23.9M\\
    Batch size   & 32    & 32 &   32\\
    Optimizer&   SGD & SGD    & SGD \\
    Momentum & 0.9 & 0.9 & 0.9\\
    Weight decay & 0.0005 & 0.0005 & -\\
    Epochs (Steps)    & 200 & 200 &  10000 (Steps)\\
    Learning rate & 0.01 & 0.01 & 0.003 (Staircase decay)\\
    $\lambda$ (value clipping) & 2.5 & 2.5 & -\\
    \bottomrule
    \end{tabular}
\end{table}

\vspace{-1cm}
\begin{table}
    
    \centering
    \caption{Hyperparameter configurations for generators}\label{tab:eval:gen_training}
    \begin{tabular}{ @{\hspace{1em}} l @{\hspace{1em}} | @{\hspace{1em}} c @{\hspace{1em}} c @{\hspace{1em}} c @{\hspace{1em}} }
    \toprule
     Parameters & $\mathcal{G}_\text{ViT}$ & $\mathcal{G}_\text{Swin}$ & $\mathcal{G}_\text{Siamese}$\\
    \midrule
    Batch size   & 32    & 16 &   32\\
    Optimizer&   Adam& Adam    &Adam \\
    $\beta_1$ \& $\beta_2$ for Adam & 0.5 \& 0.999& 0.5 \& 0.999& 0.5 \& 0.999\\
    Magnitude of perturbations &10 & 10&  10\\
    Epochs    &200 & 200&  100\\
    Learning rate & 0.0002 & 0.0002 & 0.0002\\
    Target probability, $p_{\text{adversarial}}$ & 0.5 & 0.5 & 0.5\\
    \bottomrule
    \end{tabular}
\end{table}

\end{document}